\documentclass[journal]{IEEEtran}
\usepackage{cite}
\usepackage{amsmath,amssymb,amsfonts}
\usepackage{algorithmic}
\usepackage{booktabs}
\usepackage{tabularx}
\usepackage{float}
\usepackage{graphicx}
\usepackage{textcomp}
\usepackage{algorithm}\usepackage{multirow}
\usepackage{xcolor}

\definecolor{storedblue}{RGB}{220,235,255}
\definecolor{derivedred}{RGB}{255,230,230}
\def\BibTeX{{\rm B\kern-.05em{\sc i\kern-.025em b}\kern-.08em
    T\kern-.1667em\lower.7ex\hbox{E}\kern-.125emX}}

\bstctlcite{IEEEexample:BSTcontrol}
\usepackage{eso-pic}

\begin{document}

\AddToShipoutPictureFG*{%
  \AtPageLowerLeft{%
    \raisebox{7mm}{%
      \makebox[\paperwidth][c]{%
        \parbox{0.92\paperwidth}{%
          \centering\footnotesize
          This work has been submitted to the IEEE for possible publication.
          Copyright may be transferred without notice, after which this
          version may no longer be accessible.
        }%
      }%
    }%
  }%
}

\title{Large-Time-Step Operation in a Volume Integral Equation for Dielectric Scattering}

\author{%
Sushil~Kumar,
Giampiero~Gerini,~\IEEEmembership{Senior Member,~IEEE},
and M.~C.~van~Beurden,~\IEEEmembership{Senior Member,~IEEE}%
\thanks{Sushil Kumar and M. C. van Beurden are with the Department of
Electrical Engineering, Eindhoven University of Technology,
Eindhoven, The Netherlands
(e-mail: s.kumar2@tue.nl; m.c.v.beurden@tue.nl).
Corresponding author: Sushil Kumar.}%
\thanks{Giampiero Gerini is with the Optics Department, Netherlands
Organisation for Applied Scientific Research (TNO), Delft,
The Netherlands, and also with the Department of Electrical Engineering,
Eindhoven University of Technology, Eindhoven, The Netherlands
(e-mail: giampiero.gerini@tno.nl).}%
}

\maketitle

\begin{abstract}
In transient electromagnetic analysis, explicit time-domain solvers are restricted by the Courant--Friedrichs--Lewy (CFL) condition, making finely discretized dielectric scattering problems computationally expensive. This work investigates large-time-step operation in a marching-on-in-time time-domain current-density volume integral equation (MOT-JVIE) solver for dielectric scattering. For the considered band-limited excitations, accurate transient analysis is demonstrated for time steps up to 16 times larger than the reference CFL-limited time step associated with the voxel discretization. The study reveals a fundamental computational shift in the large-time-step regime. As the time-step size increases, the present-time causal interaction region expands, increasing the number of nonzero entries in the present-time interaction matrix and causing the dominant computational cost to transition from history-term evaluations to repeated matrix--vector products involving this matrix. Consequently, the present-time interaction matrix emerges as the principal scalability bottleneck in the large-time-step regime. To address this bottleneck, a matrix-free FFT-based matrix--vector-product strategy that exploits the multilevel Toeplitz structure of the Green-function-related volume-integral operator is employed for the present-time interaction matrix. The proposed framework is evaluated through an inhomogeneous dielectric cube and an $8 \times 8$ array of inhomogeneous dielectric nanopillars representative of multiscale metasurface structures, demonstrating more than an order-of-magnitude reduction in computational cost. In single-threaded execution, the method is demonstrated for 15.6 million unknowns, providing a large-scale MOT-JVIE demonstration beyond 15 million unknowns on one CPU thread.
\end{abstract}



\section{Introduction}\label{sec:introduction}

Time-domain electromagnetic analysis is essential for broadband excitation, transient scattering, and coupled multiphysics problems, where the field evolution must be resolved directly rather than reconstructed from repeated frequency-domain solves \cite{ergul2019new,jin2019multiphysics,zhang2020electromagnetic}. Explicit differential-equation solvers such as finite-difference time-domain (FDTD) are the de-facto standard, but their time-step size is constrained by the Courant–Friedrichs–Lewy (CFL) condition imposed by the spatial discretization \cite{Taflove1995}. This constraint is particularly restrictive for multiscale metamaterial and nanophotonic structures, where electrically large domains contain deeply subwavelength features, thin dielectric regions, and abrupt material discontinuities. Resolving these features requires a fine spatial discretization, which imposes a small CFL-limited time step and, consequently, a prohibitively large number of time steps. Advances in three-dimensional dielectric fabrication further strengthen the need for scalable time-domain solvers for this class of problems~\cite{Hu2024LaserInduced,Dorrah2025FreeStanding}. Unlike explicit partial differential equation (PDE) based solvers, Marching-on-in-Time (MOT) based time-domain current density volume integral equation (TDJVIE) solvers are not inherently governed by the CFL stability limit because the solution is not advanced through local time updates of Maxwell’s differential equations. Nevertheless, stability, conditioning, and temporal discretization accuracy remain important considerations. 

In this work, the term large time step refers to temporal discretizations satisfying \(\Delta t > \Delta t_{\mathrm{CFL}}\), where \(\Delta t_{\mathrm{CFL}}\) denotes the CFL-limited time step associated with the voxel discretization. The quantity \(\Delta t_{\mathrm{CFL}}\) is used only as a reference temporal scale and does not represent a stability limit of the TDJVIE solver. Large-time-step operation has been studied in PDE-based and surface time-domain integral-equation (TDIE) solvers. Stabilized time-domain finite-element methods have been developed for arbitrarily large time steps \cite{He2012LargeStepTDFEM,Makwana2018LargeStepsEM,Makwana2019LargeStepsLayered}, and explicit unconditionally stable FDTD schemes have also been proposed \cite{Gaffar2014USFDTD}. For surface TDIE formulations, time-domain-electric-field-integral-equation (TD-EFIE) low-frequency breakdown has been analyzed \cite{Andriulli2009TDEFIE}, while DC-stable large-time-step TD-EFIE formulations based on quasi-Helmholtz projectors have been introduced \cite{Beghein2015LargeStepTDEFIE}. More recently, large-time-step implicit Runge--Kutta TD-EFIE formulations have been developed \cite{Dely2020IRKTDEFIE}. In contrast, the present work focuses on computational efficiency in large-time-step dielectric TDJVIE analysis. Unlike prior large-time-step studies that primarily emphasize stability and CFL-independent formulations, the present work treats the time-step size as a controllable parameter for optimizing computational performance in MOT-JVIE analysis. 

For penetrable dielectric structures, time-domain volume integral equations (TDVIEs) provide an attractive formulation because the radiation conditions are inherently satisfied and the unknowns are confined to material regions that differ from the background medium, thereby avoiding discretization of the surrounding free space \cite{ergul2019new,Sankaran2019RightTools,deHoop2008Reciprocity}. For dielectric scattering problems, the TDJVIE formulation represents the induced polarization response through volumetric contrast currents while accommodating material inhomogeneity. Existing TDVIE research has investigated dielectric formulations, discretization strategies, stability, and fast solution techniques \cite{Gres2001TransientVIE,Shanker2004FastLossy,Kobidze2005FastTDIEScheme,Hu2016NonconformalTDVIE,Cao2017HighOrderNystromTDVIE,Sayed2015,Sayed2020ExplicitMFVIE}. Among these approaches, marching-on-in-time (MOT) schemes are used because they preserve causality through temporal marching, while systematically incorporating history interactions from preceding time steps. The computational cost of MOT-based TDJVIE solvers is therefore strongly influenced by the temporal marching structure, particularly through the repeated evaluation of present-time and history interaction terms~\cite{VanDiepen2024PIERB104}.

The objective of this work is to determine how large-time-step operation changes the computational structure of the MOT-TDJVIE solver for dielectric scattering, provided that the time step remains admissible with respect to the temporal sampling requirement. This question is first examined for a homogeneous dielectric cube, which provides a controlled setting for assessing solver behavior and transient accuracy as the time step exceeds the reference CFL scale. The computational cost is decomposed into two primary components: the cost of constructing the excitation vector from current densities at preceding time steps and the cost of solving the linear system at each time step. This decomposition clarifies the tradeoff between the reduced cost of temporal marching and history evaluation and the increased cost associated with an increasingly present-time matrix and consequently computationally more expensive matrix-vector products for larger time steps.

This work advances large-scale transient analysis of electrically large, inhomogeneous dielectric media. The volume-integral-equation formulation directly captures spatially varying material properties and internal electromagnetic fields while restricting discretization to the material volume. The present-time interaction matrix is identified as the dominant large-time-step bottleneck, shifting the acceleration target from history-term evaluations to matrix--vector products involving this matrix. An established FFT framework is applied to the Green-function part of the present-time interaction matrix, enabling a matrix-free treatment demonstrated up to 15.6 million unknowns in single-threaded execution. To the best of the authors' knowledge, this is the first reported single-threaded large-time-step MOT-JVIE demonstration beyond 15 million unknowns. The remainder of the paper is organized as follows. Section~\ref{Sec:MOT_TDJVIE} presents the MOT-based TDJVIE formulation. Section~\ref{Sec:Large_time_step_tradeoff} analyzes the computational behavior of large-time-step MOT-JVIE operation. Section~\ref{Sec:FFt_accelerated_Dense_MVP} presents matrix-free FFT-Based matrix-vector product (MVP) for the present-time interaction matrix, and Section \ref{Sec:results} demonstrates the large-time-step MOT-JVIE framework on representative dielectric scattering problems. Finally, Section~\ref{sec:Conclusion} concludes the paper.


\section{MOT-Based TDJVIE Formulation} \label{Sec:MOT_TDJVIE}

We first summarize the MOT-based TDJVIE formulation and space--time discretization discussed in~\cite{VanDiepen2024PIERB104} and used throughout the paper. Subsequently, conditions for the selection of large time steps for accurate transient analysis are also discussed. A dielectric object embedded in free space occupies the volume $V_{\varepsilon}$ and is described by the permittivity function $\varepsilon(\mathbf{r})$, with $\mathbf{r}$ the position vector. The time-domain current-density volume integral equation for scattering by this dielectric object is given by
\begin{multline}
\left(
\varepsilon_r(\mathbf r)-1
\right)
\varepsilon_0
\frac{\partial}{\partial t}
\mathbf E^{i}(\mathbf r,t)
=
\varepsilon_r(\mathbf r)
\mathbf J_{\varepsilon}(\mathbf r,t)
\\
-
\left(
\varepsilon_r(\mathbf r)-1
\right)
\nabla\times\nabla\times
\iiint_{V_{\varepsilon}}
\frac{
\mathbf J_{\varepsilon}(\mathbf r',\tau_0)
}{
4\pi R
}
\,dV',
\label{eq:TDJVIE}
\end{multline}
where the induced contrast current density  $\mathbf{J}_{\varepsilon}$ is related to the total electric field $\mathbf{E}$ through
\begin{equation}
\mathbf{J}_{\varepsilon}(\mathbf r,t)
=
\left(
\varepsilon(\mathbf r)-\varepsilon_0
\right)
\frac{\partial}{\partial t}
\mathbf{E}(\mathbf r,t).
\label{eq:Contrast_current_density}
\end{equation}
Further, \(\mathbf{E(\mathbf r,t)} = \mathbf {E^{i}(\mathbf r,t)} + \mathbf{E_s(\mathbf r,t)}\), with \(\mathbf E^{i}\) and \(\mathbf E_s\) the incident and scattered electric fields, respectively. Furthermore,
\[
R=\|\mathbf r-\mathbf r'\|,
\qquad
\tau_0=t-\frac{R}{c_0},
\]
where \(\mathbf r\) and \(\mathbf r'\) are the observation and source coordinates, respectively, and \(c_0\) is the speed of light in free space. The relative permittivity is defined as \(\varepsilon_r(\mathbf r)=\varepsilon(\mathbf r)/\varepsilon_0\). The induced contrast current density vanishes automatically in regions where \(\varepsilon_r(\mathbf r)=1\).

The TDJVIE is discretized using a separable space--time expansion of the contrast current density. A Dirac-delta temporal testing procedure enforces the equation at uniformly sampled time instants, avoiding explicit temporal integration during testing. Combined with causal temporal basis functions, this yields an implicit marching-on-in-time (MOT) formulation in which the solution at each time step depends only on previously computed states. Spatial discretization is performed on a uniform Cartesian voxel grid using piecewise-constant basis and testing functions for each Cartesian component of the contrast current density, while quadratic-spline temporal basis functions of polynomial order \(p=2\) are employed to maintain stability and robustness of the MOT discretization~\cite{VanDiepen2024PIERB104}.

Application of the space--time discretization to the TDJVIE yields the following implicit causal marching-on-in-time (MOT) system:

\begin{equation}
\mathbf{Z}_0\mathbf{J}_n
=
\mathbf{E}^{i}_n
-
\sum_{n'=n-1}^{n-\ell}
\mathbf{Z}_{n-n'}
\mathbf{J}_{n'},
\label{eq:mot_tdjvie}
\end{equation}
where \(\mathbf{E}^{i}_n\) is the discretized incident-field excitation vector at the \(n\)-th time step, \(\mathbf{J}_n\) denotes the vector of unknown contrast-current expansion coefficients at the \(n\)-th time step. The matrix \(\mathbf{Z}_0\) is the present-time interaction matrix, while \(\mathbf{Z}_{n-n'}\) represents past-time or history interaction matrices associated with fields generated at preceding time steps. Consequently, computation of \(\mathbf{J}_n\) requires repeated iterative solution of the linear system associated with the present-time interaction matrix \(\mathbf{Z}_0\). 

The interaction matrices are precomputed for \(n-n'=0,\ldots,\ell\), where the maximum temporal interaction lag is
\begin{equation}
\ell
=
\left\lceil
\frac{R_{\max}}{c_0\Delta t}
\right\rceil
+
p,
\label{eq:lmax}
\end{equation}
with \(R_{\max}\) denoting the maximum source--observer distance within the computational domain. In the present work, the linear system associated with the present-time interaction matrix \(\mathbf{Z}_0\) is solved at each time step using the transpose-free quasi-minimal residual (TFQMR) iterative method~\cite{Freund1993TFQMR}. No additional regularization or stabilization is employed in the present large-time-step analysis. In the notation of the \(\delta\)-regularized MOT formulation pertaining to Eq.~(35) in \cite{VanDiepen2024PIERB106}, where the interaction matrices are modified as \(\mathbf{Z}^{\delta}_{n}=\mathbf{Z}_{n}+\delta_{n}\mathbf{I}\), the present study corresponds to setting \(\delta_{n}=0\) for all time indices. Hence, \(\mathbf{Z}^{\delta}_{n}=\mathbf{Z}_{n}\).

In the MOT-based TDJVIE solver, the time step \(\Delta t\) is selected primarily from a temporal sampling requirement rather than a CFL stability constraint. If the incident and scattered fields are effectively band-limited to a frequency \(f_{\max}\), then the Nyquist-Shannon sampling theorem implies the alias-free temporal sampling condition~\cite{Shannon1949,Unser2000Sampling}
\begin{IEEEeqnarray}{rCl}
\Delta t \;\le\; \frac{1}{2\,\kappa\,f_{\max}},
\label{eq:dt_nyquist_tdjevie}
\end{IEEEeqnarray}
where \(\kappa \ge 1\) is an oversampling factor used to preserve temporal accuracy. Consequently, low-frequency excitations permit larger time steps while maintaining accurate transient-field reconstruction. Accordingly, the present work investigates whether such temporally admissible large time steps can also provide computational advantages in MOT-TDJVIE analysis.


\section{Large-Time-Step Computational Tradeoffs} \label{Sec:Large_time_step_tradeoff}


\subsection{Bottleneck transition at large time steps} \label{Subsec:bottleneck_large_time_step}

In MOT-based TDJVIE analysis, increasing the time-step size \(\Delta t\) reduces the number of time steps required to reach a prescribed final time, but simultaneously alters the computational work performed at each step. Unlike explicit differential-equation-based solvers, MOT-based formulations do not seem to be governed by a CFL condition associated with the spatial discretization. Consequently, time steps larger than the equivalent CFL limit can be employed, provided that the transient response remains temporally resolved according to the sampling requirement. Large time steps are particularly attractive when the objective is to evaluate the transient response at a prescribed final time or at sparsely sampled observation times rather than to resolve every intermediate temporal state. Reducing the number of time steps decreases the amount of intermediate information that must be computed and stored, which is especially beneficial in large-scale TDIE simulations where each time step requires nonlocal operator evaluations and iterative linear solves.

\begin{figure}[htb]
    \centering
    \includegraphics[width=1\columnwidth]{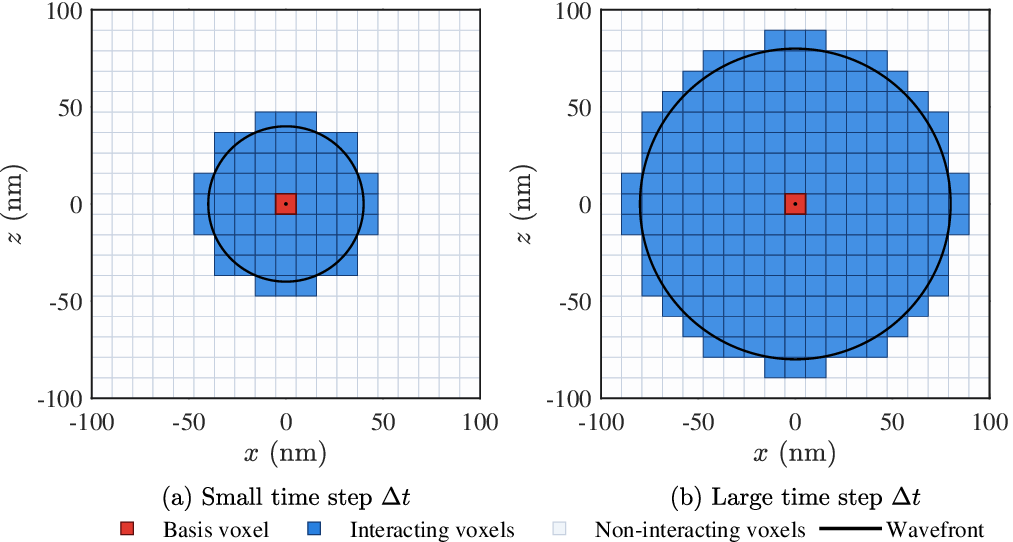}
    \caption{Effect of the time-step size on the voxel-interaction pattern and the associated computational bottleneck for a \(200\,\mathrm{nm}\times200\,\mathrm{nm}\times200\,\mathrm{nm}\) dielectric cube shown on the \(x\)-\(z\) mid-plane. (a) For a small \(\Delta t\), the spherical wavefront intersects or encloses only nearby voxels around the basis voxel, producing a sparse present-time interaction. (b) For a larger \(\Delta t\), the wavefront encloses a larger portion of the domain, increasing the number of interacting voxels and the corresponding number of nonzero entries in \(\mathbf Z_0\).}
    \label{fig:bottleneck_transition}
\end{figure}

To illustrate how the interaction behavior changes with the time-step size, consider a \(200\,\mathrm{nm}\times200\,\mathrm{nm}\times200\,\mathrm{nm}\) dielectric cube discretized into uniform voxels. Fig.~\ref{fig:bottleneck_transition} provides a geometric interpretation of how the sparsity of the present-time matrix \(\mathbf Z_0\) depends on \(\Delta t\). The figure is shown in the 2-D \(x\)-\(z\) mid-plane for clarity, whereas the actual TDVIE discretization is three-dimensional. Therefore, the circular wavefronts in the figure should be interpreted as spherical wavefronts centered at the associated red basis voxel in the 3-D voxelized domain. For a fixed basis voxel, the interaction matrices are populated by the testing voxels intersected or enclosed by the spherical wavefront associated with the propagation distance over the corresponding time interval. For \(\mathbf Z_0\), these testing voxels are shown in blue; hence, the number of blue voxels directly represents the number of nonzero entries in the corresponding column of \(\mathbf Z_0\), apart from a factor of three for each of the Cartesian directions. To analyze how the computational cost changes with increasing time-step size, the MOT system in \eqref{eq:mot_tdjvie} is decomposed into its left-hand-side (LHS) and right-hand-side (RHS) components:
\begin{equation}
\mathrm{LHS} = \mathbf{Z}_0 \mathbf{J}_n, 
\quad
\mathrm{RHS} = \mathbf{E}_n^{i} - \sum_{n'=n-1}^{n-\ell} \mathbf{Z}_{n-n'} \mathbf{J}_{n'}
\label{eq:lhs_rhs_def}
\end{equation}

\subsubsection{Small time-step regime}
\label{subsubsec:small_time_steps}
For small time-step choices, as illustrated in Fig.~\ref{fig:bottleneck_transition}(a), the electromagnetic wavefront advances only over a limited spatial region during one time step. Consequently, only a small neighborhood of testing voxels contributes to the column of \(\mathbf Z_0\) associated with the red basis voxel, yielding a sparse present-time matrix. The delayed-history terms follow the same geometric construction through subsequent spherical shells separated by \(\Delta t\). Since the wavefront requires many marching intervals to traverse the computational domain, the number of delayed-history terms increases according to Eq.~\eqref{eq:lmax}. Thus, although the individual interaction matrices remain sparse, the delayed-history contribution becomes computationally costly due to the accumulation of many terms during the MOT procedure. Direct evaluation of these terms can therefore dominate the overall runtime. Existing fast algorithms, including plane-wave time-domain (PWTD) methods~\cite{Ergin1999,Liu2016PWTDAcceleratedExplicitEFVIE} and FFT-based acceleration techniques~\cite{Yilmaz2004,VanDiepen2024PIERB106}, primarily target this bottleneck by exploiting the structured nature of the history interaction matrices and reducing the delayed-history evaluation cost to approximately \(\mathcal{O}(N\log N)\). In this regime, the matrix--vector product with \(\mathbf Z_0\), employed in the TFQMR iterative solver, remains relatively inexpensive.

\subsubsection{Large time-step regime}
\label{subsubsec:large_time_steps}
For large time-step choices, as illustrated in Fig.~\ref{fig:bottleneck_transition}(b), the wavefront spans a much larger portion of the voxelized domain during a single time step. Following the same interpretation as above, many more blue voxels are intersected or enclosed by the spherical wavefront, and hence many more nonzero entries are generated in the column of \(\mathbf Z_0\) associated with the red basis voxel. The present-time matrix \(\mathbf{Z}_0\) is stored in MATLAB sparse format. However, the number of nonzero entries in \(\mathbf{Z}_0\) increases as \(\Delta t\) increases. Accordingly, \(\mathbf{Z}_0\) becomes progressively less sparse for larger time steps. The matrix approaches a densely populated form only in the limiting case where the present-time wavefront covers most of the computational domain within a single time step. The delayed-history terms are populated in an analogous manner by subsequent spherical shells separated in time by \(\Delta t\). These terms continue to contribute until the wavefront has traversed the entire computational domain, with the required number of terms determined by Eq.~\eqref{eq:lmax}. For larger \(\Delta t\), fewer delayed-history terms are needed; however, each term involves a larger set of voxel interactions. Consequently, the cost of applying each interaction matrix increases. In particular, the matrix--vector product with \(\mathbf Z_0\) becomes substantially more expensive because it is performed at every TFQMR iteration and at every MOT step. Hence, as the time-step size increases, the dominant computational bottleneck shifts from the accumulation of many sparse delayed-history terms to the repeated evaluation of the present-time matrix \(\mathbf Z_0\) in Eq.~\eqref{eq:mot_tdjvie}.

Overall, Fig.~\ref{fig:bottleneck_transition} illustrates the time-step-dependent transition of the dominant computational bottleneck. Throughout the numerical studies presented here, the RHS contribution in Eq.~\eqref{eq:lhs_rhs_def} is accelerated using the FFT-based delayed-history scheme in~\cite{VanDiepen2024PIERB106}. With the RHS cost reduced, increasing \(\Delta t\) shifts the bottleneck toward the present-time matrix \(\mathbf Z_0\), whose matrix--vector product is required at every TFQMR iteration and each time step. This motivates an acceleration strategy specifically targeting \(\mathbf Z_0\) in the large-time-step regime.

\subsection{Homogeneous-Dielectric-Cube Analysis} \label{subsec:Homogenous_Cube_analysis}
\begin{figure}[htb]
    \centering
\includegraphics[width=\columnwidth]{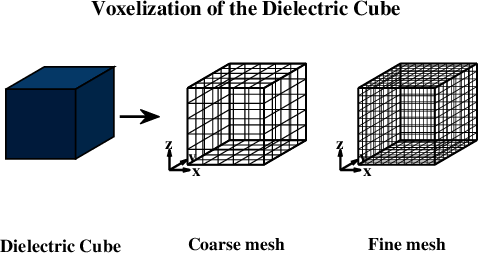}
    \caption{Voxelization of a homogeneous dielectric cube into coarse and fine meshes. In reality, the coarse and fine discretizations consist of $40 \times 40 \times 40$ and $80 \times 80 \times 80$ voxels, respectively. The grid lines shown here are only for the purpose of illustration and do not correspond to the actual number of voxels.}
    \label{fig:Dielectric_cube_voxelization}
\end{figure}

To quantitatively investigate the computational tradeoffs associated with enlarged time steps discussed in Section~\ref{Subsec:bottleneck_large_time_step}, we consider a homogeneous dielectric cube with edge length \(200~\mathrm{nm}\) and relative permittivity \(\varepsilon_r=12\), as shown in Fig.~\ref{fig:Dielectric_cube_voxelization}. A coarse and a fine isotropic voxel discretization, see Table~\ref{tab:voxelization},  are employed to evaluate the combined influence of spatial resolution and time-step scaling on the computational behavior of the MOT-JVIE formulation.
\begin{table}[tb]
\centering
\small
\caption{Voxel-grid discretizations of the  $200~\mathrm{nm^3}$ cube \\
(Voxel dimensions)}
\label{tab:voxelization}
\setlength{\tabcolsep}{3pt} 
\centering
\begin{tabular}{lll}
\toprule
\textbf{Mesh} & $\mathbf{N_x \times N_y \times N_z}$ & $(\Delta x,\Delta y,\Delta z)$ [nm] \\
\midrule
Coarse & $40 \times 40 \times 40$   & $(5.0,\,5.0,\,5.0)$ \\
Fine   & $80 \times 80 \times 80$ & $(2.5,\,2.5,\,2.5)$ \\
\bottomrule
\end{tabular}
\end{table}

The cube is illuminated by an \(x\)-polarized modulated Gaussian plane wave propagating along \(-\hat{\mathbf z}\) given by
\begin{equation}
\mathbf{E}^{\mathrm{inc}}(\mathbf{r},t)
= E_0\,\hat{\mathbf{x}}\, \frac{1}{\sigma\sqrt{2\pi}}\exp\!\left(-\frac{\tau^2}{2\sigma^2}\right)\cos\!\big(2\pi f_0 \tau\big),
\label{eq:Einc_modGauss}
\end{equation}
where
\begin{equation}
\tau=t-t_0-\frac{\hat{\mathbf{k}}\cdot\mathbf{r}}{c_0},
\ \ \hat{\mathbf{k}}=-\hat{\mathbf{z}}.
\label{eq:Einc_tau_factor}
\end{equation}

\begin{table}[tb]
\centering
\small
\caption{Parameters of the incident plane wave}
\label{tab:excitation_parameters}
\setlength{\tabcolsep}{3pt} 
\centering
\begin{tabular}{llll}
\toprule
\textbf{Symbol} & \textbf{Description} & \textbf{Value} & \textbf{Unit} \\
\hline
$E_0$   & Peak electric-field amplitude & $1$      & $\mathrm{V/m}$ \\
$\sigma$ & Pulse width                    & $0.8$   & $\mathrm{fs}$ \\
$t_0$   & Initial pulse-center delay      & $4.8$  & $\mathrm{fs}$ \\
$f_0$   & Center frequency                & $1.537$  & $\mathrm{PHz}$ \\
\hline
\end{tabular}
\end{table}

The parameters of the incident wave are given in Table~\ref{tab:excitation_parameters}. This excitation is broadband, with significant spectral content spanning the free-space wavelength range of approximately \(180\)--\(210~\mathrm{nm}\), thereby enabling assessment of time-step sensitivity over a wide frequency range. 

To investigate enlarged time-step behavior, the MOT-JVIE formulation is evaluated using
\begin{equation}
\Delta t = \alpha \Delta t_{\mathrm{CFL}},
\quad
\alpha \in \{1,2,4,8,16,32\},
\label{eq:dt_set}
\end{equation}
where \(\Delta t_{\mathrm{CFL}} = 0.005\)~fs denotes the reference temporal discretization associated with the fine voxelization. This value is used for both coarse and fine voxelizations. For \(\lambda_{\min}=180~\mathrm{nm}\), Eq.~\eqref{eq:dt_nyquist_tdjevie} with \(\kappa=2\) gives \(\Delta t_{\mathrm{Nyq}}\approx 0.150~\mathrm{fs}\), corresponding to \(\alpha_{\max}\approx 30\). Thus, the enlarged time-step cases \(\alpha=1,2,4,8,\) and \(16\) satisfy the Nyquist limit, whereas \(\alpha=32\) exceeds it. This configuration enables direct evaluation of the accuracy and computational tradeoffs induced by increasing \(\Delta t\).

\begin{figure}[tb]
\centering
\includegraphics[width=0.35\textwidth]{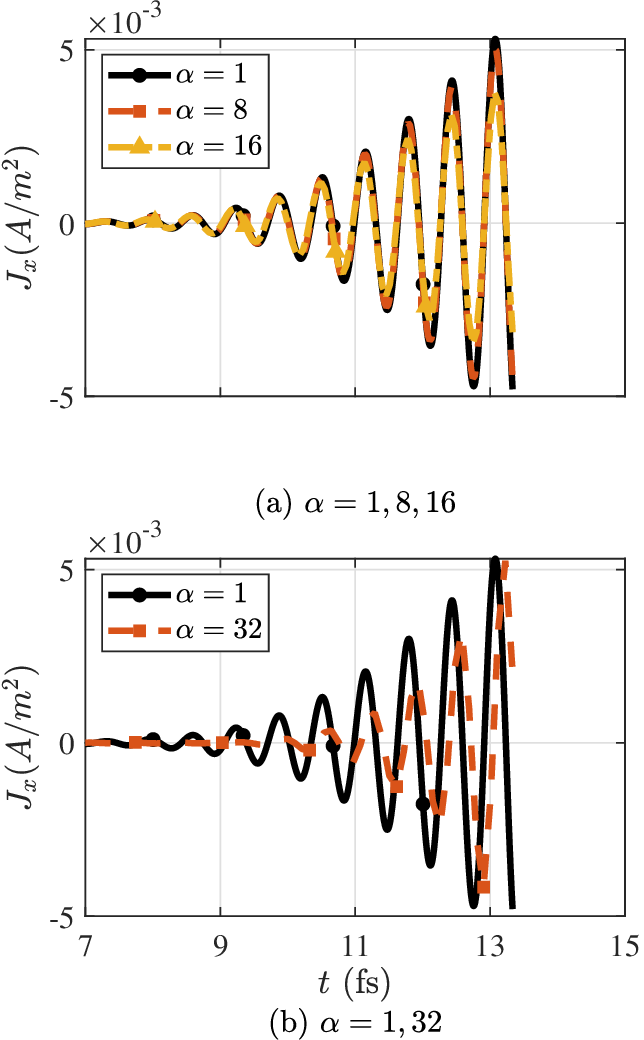}
\caption{Time-domain responses of the $x$-directed contrast current density component \(J_x\) for different time-step scaling factors \(\alpha\) in a homogeneous cube discretized with fine voxelization. The \(\alpha=1\) solution is used as the reference. In (a), the responses for \(\alpha=8\) and \(\alpha=16\) follow the same wave behavior with the reference over the plotted time window. In (b), the larger time-step choice \(\alpha=32\) exhibits noticeable amplitude and phase deviations.}
\label{fig:J_components_total_time_comparison}
\end{figure}

Before analyzing computational performance, the effect of enlarged time steps on temporal accuracy is examined using the transient contrast current density. Figure~\ref{fig:J_components_total_time_comparison} shows the \(J_x\) response, i.e. the $x$ component of the contrast current density, evaluated at the voxel (41,41,41) on the fine voxelization, located close to the center of the dielectric cube, for different time-step scaling factors. As shown in Fig.~\ref{fig:J_components_total_time_comparison}(a), the responses for \(\alpha=8\) and \(\alpha=16\) closely follow the reference solution obtained with \(\alpha=1\), confirming accurate large-time-step marching beyond the CFL-limited step size. In contrast, Fig.~\ref{fig:J_components_total_time_comparison}(b) shows that \(\alpha=32\) introduces noticeable amplitude and phase deviations, indicating that the solution accuracy deteriorates once the time step exceeds the Nyquist criterion.

\begin{table}[htb]
\centering
\caption{Sparsity of $\mathbf{Z}_0$ with time-step scaling factor $\alpha$.}
\label{tab:sparsity_singlecol}

\renewcommand{\arraystretch}{0.75}
\setlength{\tabcolsep}{3.5pt}
\footnotesize

\begin{tabular}{c c r c c}
\hline
\textbf{Grid} &
$\boldsymbol{\alpha}$ &
\textbf{$N_{\mathrm{NZ}}$ $(\times 10^{6})$} &
\textbf{Fill (\%)} &
\textbf{$N_{\mathrm{NZ}}(\alpha)/N_{\mathrm{NZ}}(1)$} \\
\hline

$40^3$ & 1  & 2.37   & 0.0064  & 1.00 \\
       & 2  & 9.24   & 0.0251  & 3.90 \\
       & 4  & 9.24   & 0.0251  & 3.90 \\
       & 8  & 45.56  & 0.1236  & 19.24 \\
       & 16 & 171.48 & 0.4652  & 72.40 \\
       & 32 & 793.95 & 2.1537  & 335.23 \\

\hline

$80^3$ & 1  & 16.91    & 0.00072 & 1.00 \\
       & 2  & 76.09    & 0.00323 & 4.50 \\
       & 4  & 382.40   & 0.0162  & 22.61 \\
       & 8  & 1470.00  & 0.0623  & 86.93 \\
       & 16 & 7168.80  & 0.3039  & 423.95 \\
       & 32 & 29100.00 & 1.2334  & 1720.93 \\

\hline

\end{tabular}

\vspace{2pt}
\centering
\footnotesize{
$N_{\mathrm{NZ}}$ = number of non-zero entries; 
 \\ Fill (\%) = percentage of non-zero entries relative to total matrix size; \\ $N_{\mathrm{NZ}}(\alpha)/N_{\mathrm{NZ}}(1)$ = ratio of nonzero entries relative to the $\alpha=1$ case.}
\end{table}

\begin{figure}[htb]
    \centering
\includegraphics[width=0.45\textwidth]{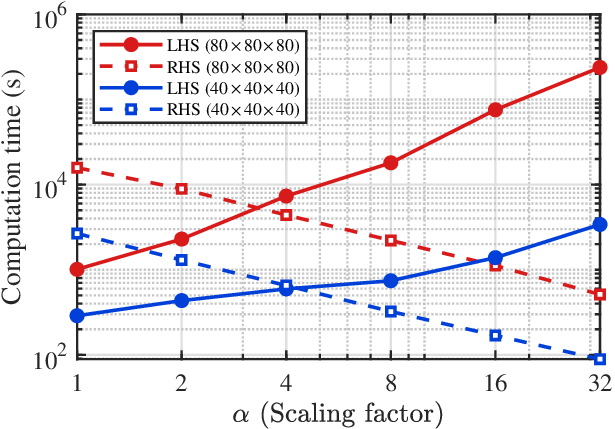}
    \caption{Comparison of the computation times associated with the LHS and RHS terms, defined in \eqref{eq:lhs_rhs_def}, plotted versus time-step scaling factor $\alpha$ for the coarse and fine voxelizations shown in Table~\ref{tab:voxelization}.} \label{fig:Computation_time_LHS_RHS_Cube}
\end{figure}

\begin{figure}[tb]
    \centering
    \includegraphics[width=0.5\textwidth]{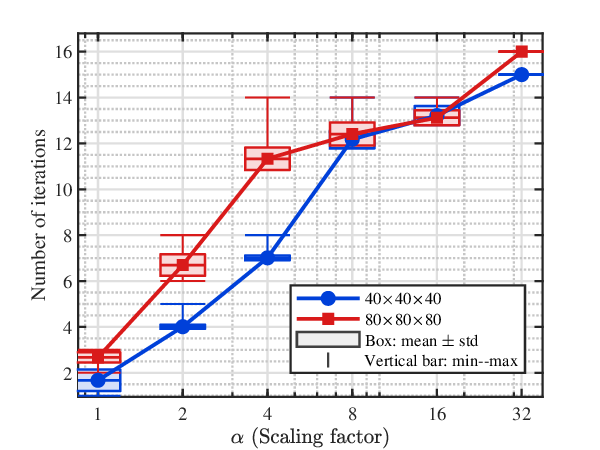}
      \caption{TFQMR iteration counts required for convergence as a function of the time-step scaling factor $\alpha$ for coarse ($40\times40\times40$) and fine ($80\times80\times80$) voxelizations.}
\label{fig:Number_of_iterations_cube_comparison}
\end{figure}

As discussed in Section~\ref{subsubsec:large_time_steps}, increasing \(\Delta t\) enlarges the interaction region within a single time step, thereby increasing the density of \(\mathbf Z_0\). Table~\ref{tab:sparsity_singlecol} summarizes the resulting decrease in matrix sparsity for both voxel discretizations, where the matrix \(\mathbf{Z}_0\) is stored in MATLAB sparse format. Figure~\ref{fig:Computation_time_LHS_RHS_Cube} shows the computation-time breakdown associated with the LHS and RHS terms in \eqref{eq:lhs_rhs_def}. For small time-step choices, the right-hand-side (RHS) history evaluation dominates the computation time, as discussed in~\ref{subsubsec:small_time_steps}. Therefore, we use the FFT-based RHS acceleration presented in~\cite{VanDiepen2024PIERB106} for the delayed-history evaluation. However, as \(\alpha\) increases, the LHS rapidly becomes dominant due to the increased cost of applying the matrix--vector product with a  \(\mathbf Z_0\), as discussed in Section~\ref{subsubsec:large_time_steps}. This trend is more pronounced for the finer mesh, where the number of nonzero elements in $\mathbf{Z}_0$ increases substantially faster. To determine whether the increase in LHS cost originates from more iterations in the iterative solver TFQMR or from more expensive matrix--vector products, the TFQMR iteration counts are shown in Fig.~\ref{fig:Number_of_iterations_cube_comparison}. The iteration count remains low and relatively stable across both time-step and mesh variations, indicating that the dominant runtime increase does not originate from a large number of iterations and a consequent large number of matrix--vector products involving $\mathbf{Z}_0$. Instead, the computational bottleneck is primarily caused by the increased cost of applying the  present-time interaction matrix \(\mathbf Z_0\) in each TFQMR iteration.

Overall, the results confirm the computational tradeoff associated with enlarged time steps in the MOT-JVIE formulation. Although increasing \(\Delta t\) reduces the number of time steps and delayed-history evaluations, it also causes substantial densification of the present-time interaction matrix \(\mathbf Z_0\), making the repeated matrix--vector products with $\mathbf{Z}_0$ to solve  the linear system iteratively the dominant computational bottleneck. Therefore, Section~\ref{Sec:FFt_accelerated_Dense_MVP} applies a matrix-free FFT-based evaluation of the present-time interaction matrix \(\mathbf{Z}_0\) to address this large-time-step computational bottleneck.


\subsection{Validation of Large Time-Step MOT-JVIE Analysis} \label{subsec:validation_large_time_steps}

To validate the proposed large-time-step MOT-JVIE analysis against established reference solutions, the benchmark dielectric-cube problem reported in~\cite{Shi2011} is considered. The configuration consists of a dielectric cube with edge length 0.2 meter and with relative permittivity \(\varepsilon_r=3.2\) illuminated by a Gaussian plane-wave excitation with pulse width of 4 lm and $t_0$ = 6.1 lm, consistent with the setup employed in \cite{VanDiepen2024PIERB104}.

In contrast to the comparatively coarse discretization used in~\cite{VanDiepen2024PIERB104}, the present study employs a refined \(20\times20\times20\) voxel discretization to primarily isolate the influence of temporal discretization on the MOT-JVIE solution behavior. The Gaussian excitation employs pulse parameters identical to those used in \cite{Shi2011,VanDiepen2024PIERB104}, thereby enabling direct comparison with previously reported MOT-JVIE and MOD-JVIE solutions.

The induced contrast current density is sampled at the observation point
\[
\mathbf{r}=0.025\hat{x}+0.075\hat{y}+0.025\hat{z},
\]
and projected along the azimuthal direction $\hat{\phi}$ according to
\[
J_{\phi}(\mathbf{r},t)=\hat{\phi}\cdot \mathbf{J}(0.025,\,0.075,\,0.025,t),
\]
where
\[
\hat{\phi}=\frac{(-y,x,0)}{\sqrt{x^2+y^2}}.
\]

Simulations are performed for temporal discretizations ranging from \(\Delta t(\alpha =1) = 0.01\) light meter \(\mathrm{(lm)}\) to \(\Delta t(\alpha = 16) = 0.16~\mathrm{lm}\). The case \(\alpha =1\) corresponds to the reference MOT-JVIE solution reported in \cite{VanDiepen2024PIERB104}, whereas the remaining cases represent enlarged temporal discretization configurations.

\begin{figure}[b]
    \centering
    \includegraphics[width=0.45\textwidth]{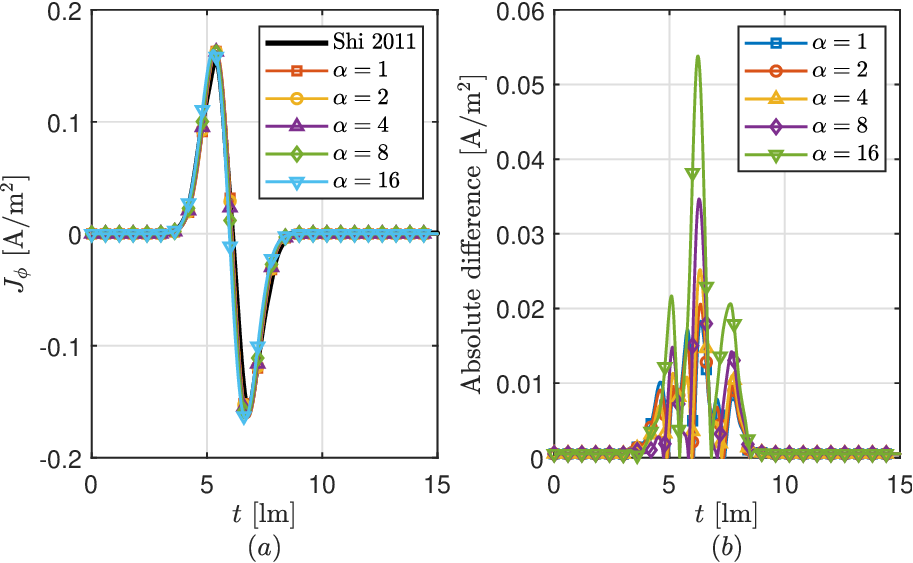}
    \caption{(a) Contrast current density responses obtained using the MOT-JVIE solver for temporal discretizations ranging from \(\alpha = 1\) to \(\alpha = 16\), compared with the reference MOD-JVIE solution in \cite{Shi2011}. The \(\alpha = 1\) case corresponds to the reference MOT-JVIE solution in \cite{VanDiepen2024PIERB104}. The current density is evaluated inside a dielectric cube with \(\varepsilon_r = 3.2\) at the observation point \((0.025,\,0.075,\,0.025)\). (b) Absolute error of the corresponding MOT-JVIE solutions relative to the reference solution in \cite{Shi2011}.}
    \label{fig:LargeTimeStepValidation}
\end{figure}

Figure~\ref{fig:LargeTimeStepValidation}(a) compares the MOT-JVIE contrast current density responses with the benchmark MOD-JVIE solution reported in~\cite{Shi2011}. A similar trend with the reference solution is observed throughout the simulation interval, even for \(\alpha = 16\) time-step choices. The dominant transient waveform characteristics, including the peak amplitudes and temporal oscillations, follow a similar trend. The corresponding absolute error profiles are shown in Fig.~\ref{fig:LargeTimeStepValidation}(b). Although the error increases gradually with \(\alpha\), the discrepancy remains below 0.06~A/m$^2$ over the entire simulation interval, demonstrating stable transient accuracy even with large time steps.


\section{Matrix-Free FFT-Based MVP for the Present-Time Interaction Matrix} \label{Sec:FFt_accelerated_Dense_MVP}

Fast algorithms for time-domain integral equations have commonly targeted the delayed-history or retarded-potential part of the marching system. Plane-wave time-domain methods and FFT-based schemes exploit structure in the history interaction matrices to reduce the cost of evaluating fields generated by previously computed current densities~\cite{Ergin1999,Liu2016PWTDAcceleratedExplicitEFVIE,VanDiepen2024PIERB106,Yilmaz2004,Yilmaz2005,Yilmaz2002}. This is the natural bottleneck in the small-time-step regime, where many delayed interactions must be accumulated and the present-time matrix remains relatively sparse. In the large-time-step regime considered here, however, the number of delayed-history terms decreases while the number of nonzero entries in the present-time interaction matrix \(\mathbf{Z}_0\) increases. Consequently, the dominant acceleration target shifts from history-term evaluations to matrix--vector products involving \(\mathbf{Z}_0\). The present-time matrix is decomposed as

\begin{equation}
    \mathbf{Z_0} = \mathbf{D} - \mathbf{XG} ,
\end{equation}
where \(\mathbf{D}\) and \(\mathbf{X}\) are diagonal matrices associated with the local constitutive and material-contrast terms, respectively, and $G$ denotes the volume-integral operator associated with the Green function and its spatial derivatives. Since \(\mathbf{D}\) and \(\mathbf{X}\) require only elementwise operations, the dominant nonlocal cost in applying \(\mathbf{Z}_0\) comes from the matrix--vector product involving \(\mathbf{G}\). For a uniform Cartesian voxel discretization in a homogeneous background medium, the entries of \(\mathbf{G}\) depend only on relative source--observer voxel offsets. Therefore, \(\mathbf{G}\) has a multilevel block Toeplitz structure that can be embedded into a multilevel circulant operator and applied using multidimensional FFTs~\cite{Zwamborn1991,Zwamborn1992,Polimeridis2014StableFFTJVIE,ChanOlkin1994ToeplitzBlock}. Following the standard Toeplitz/circulant FFT approach, each Cartesian current component is zero-padded onto an enlarged circulant grid, transformed to the spectral domain, multiplied by the precomputed spectral Green-function kernels, and then transformed back and truncated to the physical voxel grid. This procedure evaluates the Green-function contribution without explicitly forming the present-time interaction matrix \(\mathbf{Z}_0\), whose number of nonzero entries increases with the time-step size. The action of the Green-function operator on the current-density vector is written compactly as

\begin{equation}
\begin{aligned}
    \mathbf{(GJ_n)_p}
    &=
    T\Bigg\{
    \mathcal{F}^{-1}
    \Bigg[
    \sum_{q\in\{x,y,z\}}
    \Lambda_{pq}\,
    \mathcal{F}\!\left\{\mathbf{J^{\mathrm{ext}}_{q,n}}\right\}
    \Bigg]
    \Bigg\}, \\
    &\quad p\in\{x,y,z\}.
\end{aligned}
\end{equation}
where $\mathcal{F}\{\cdot\}$ and $\mathcal{F}^{-1}\{\cdot\}$ denote the three-dimensional FFT and inverse FFT, respectively, $\Lambda_{pq}$ contains the precomputed spectral coefficients of the circulant Green-function kernel, \(\mathbf{J^{\mathrm{ext}}_{q,n}}\) is the zero-padded $q$-directed current-density component, and $T\{\cdot\}$ denotes truncation back to the physical voxel grid. After the Green-function contribution is evaluated, the matrix--vector product with the present-time interaction matrix is assembled as
\begin{equation}
    \mathbf{Z_0J_n} = \mathbf{DJ_n} - \mathbf{X(GJ_n)}.
\end{equation}
In the direct implementation, \(\mathbf{Z}_0\) is explicitly assembled and stored in sparse format, so the storage and multiplication costs scale with the number of nonzero entries in \(\mathbf{Z}_0\). As shown in Section III, this number increases rapidly as the time-step size grows. In the accelerated implementation, \(\mathbf{Z}_0\) is not explicitly formed; only its action on a vector is evaluated through local diagonal operations and FFT-based convolution. Thus, the dominant present-time matrix--vector product in the large-time-step regime is evaluated in a matrix-free manner with $O(N\log N)$ operator-application complexity.


\section{Results} \label{Sec:results}

The accuracy, computational efficiency, and scalability of the matrix-free FFT-based matrix–vector product (MVP) evaluation for the present-time interaction in the large-time-step MOT-JVIE framework are assessed using two large, inhomogeneous dielectric scattering problems. The first example consists of an inhomogeneous dielectric cube whose eight octants are assigned distinct permittivity values. This example verifies the FFT-based present-time matrix--vector product against the direct matrix implementation using current-density comparisons, relative $\ell_2$-norm errors, runtime measurements, and TFQMR iteration counts over different time-step sizes. The second example is an $8\times8$ dielectric nanopillar array inspired by phase-change-material-based bilayer metasurfaces~\cite{Li2024_PCM_Metasurface}. This larger example demonstrates the applicability of the proposed implementation to multiscale volumetric scattering problems for which conventional MOT-JVIE simulations are impractical due to memory and runtime requirements. Since the underlying direct MOT-JVIE formulation has been validated previously, including comparison with CST for related dielectric scattering configurations~\cite{VanDiepen2024PIERB104,VanDiepen2024PIERB106}, the present results focus on the new aspects introduced here: large-time-step behavior, the resulting \(\mathbf{Z}_0\)-dominated bottleneck, and the accuracy and efficiency of the matrix-free present-time matrix evaluation relative to the direct implementation. All simulations reported here were performed on a computing server equipped with 2 AMD EPYC 7573ZX processors, \(2~\mathrm{TB}\) DDR4 memory, and \(2\times768~\mathrm{MB}\) L3 cache. MATLAB was restricted to single-threaded execution using \texttt{maxNumCompThreads(1)}, so all reported runtimes use one CPU thread for the iterative solves, FFTs, and matrix-vector products, without MPI, GPU, OpenMP, or multithreaded numerical-library acceleration.

\subsection{Inhomogeneous Dielectric Cube}

The inhomogeneous dielectric cube example verifies that the matrix-free FFT-based MVP for \(\mathbf{Z}_0\) preserves the solution behavior of the direct MOT-JVIE implementation in the large-time-step regime. The comparison is performed in terms of induced contrast current density, relative $\ell_2$-norm error, runtime, and TFQMR convergence for different time-step sizes.

\begin{figure}[htb]
    \centering
    \includegraphics[width=\columnwidth]{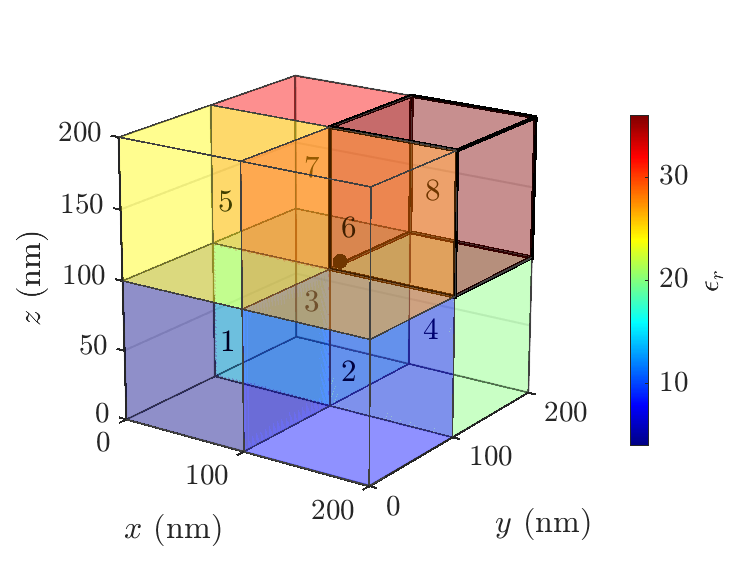}
    \caption{Octant-wise inhomogeneous dielectric cube of side length \(200~\mathrm{nm}\). The relative permittivity varies from \(\epsilon_r=4\) to \(36\) across the eight octants. The highlighted, thick-bordered octant contains the observation voxel \((41,41,41)\) near the center of the dielectric cube, on the fine \(80\times80\times80\) voxel mesh. This voxel is used to compare the \(J_x\) response obtained from the large-time-step FFT-accelerated and direct MOT-JVIE solvers.}
    \label{fig:Section_V_Inhomo_Cube}
\end{figure}

The setup follows the homogeneous-cube example in Section~\ref{subsec:Homogenous_Cube_analysis}. The cube dimensions, voxelization, excitation parameters are the same as those listed in Tables~\ref{tab:voxelization} and~\ref{tab:excitation_parameters}. The temporal discretization is defined by Eq.~\eqref{eq:dt_set}. The parameter \(\alpha\) is chosen up to 16, which satisfies the Nyquist condition, and \(\Delta t_{\mathrm{CFL}} = 0.005~\mathrm{fs}\), determined by the fine voxelization as discussed in Section~\ref{subsec:Homogenous_Cube_analysis}. To introduce volumetric inhomogeneity, the cube is partitioned into eight octants with progressively increasing relative permittivity, as shown in Fig.~\ref{fig:Section_V_Inhomo_Cube}. The octants are assigned $\varepsilon_r \in \{4,8,16,20,24,28,32,36\}$, in the order following the numbering of the octants, which introduces sharp dielectric discontinuities and nonuniform volumetric coupling within the computational domain.

\begin{figure}[htb]
    \centering
    \includegraphics[width=0.48\textwidth]{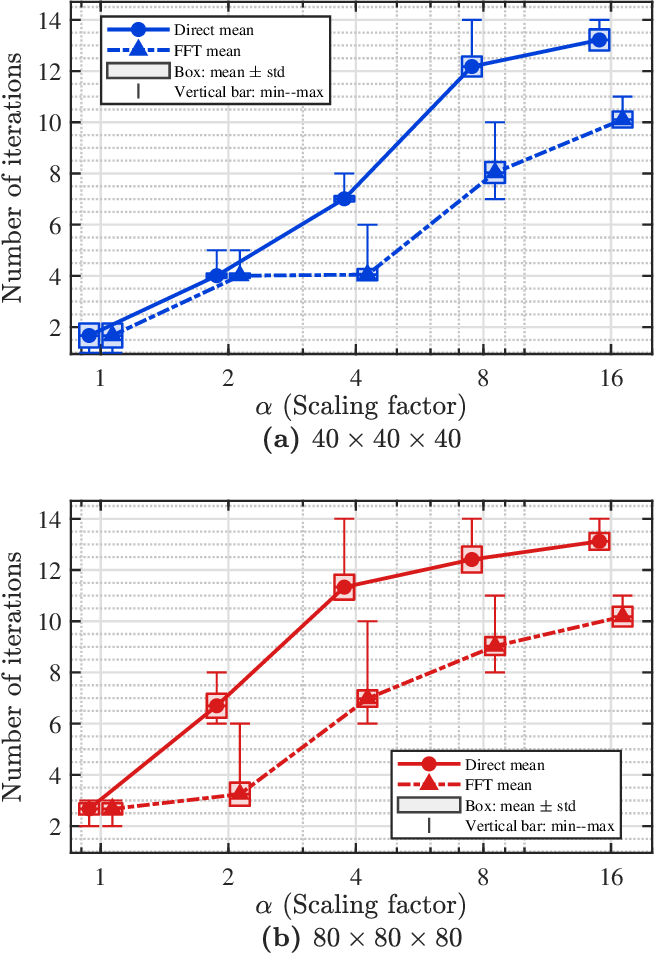}
    \caption{TFQMR iteration counts versus time-step scaling factor $\alpha$ for the direct and FFT-based MVP implementations: (a) coarse $40\times40\times40$ and (b) fine $80\times80\times80$ voxelizations.}
    \label{fig:Section_V_inhomo_Cube_No_of_iteratn_direct_fft}
\end{figure}

Fig.~\ref{fig:Section_V_inhomo_Cube_No_of_iteratn_direct_fft} compares the TFQMR iteration counts required for convergence for the direct and FFT-based \(\mathbf{Z}_0\) MVP implementations. For both voxelizations, the FFT-based \(\mathbf{Z}_0\) MVP follows the same trend as the direct implementation, indicating that the FFT-based \(\mathbf{Z}_0\) MVP implementation preserves the iterative behavior of the underlying MOT-JVIE system. The TFQMR tolerance for the FFT-based solver was set to $10^{-6}$ for both implementations. This value was chosen to balance the precision of the current density and the number of iterations. Reducing the tolerance further imposes a stricter residual criterion and brings the solution for the two implementations closer together, but it also increases the number of TFQMR iterations. For the present case, the tolerance value of $10^{-6}$ provides the required precision while maintaining a lower iteration count than the direct MOT-JVIE implementation.

\begin{figure}[htb]
    \centering
    \includegraphics[width=0.45\textwidth]{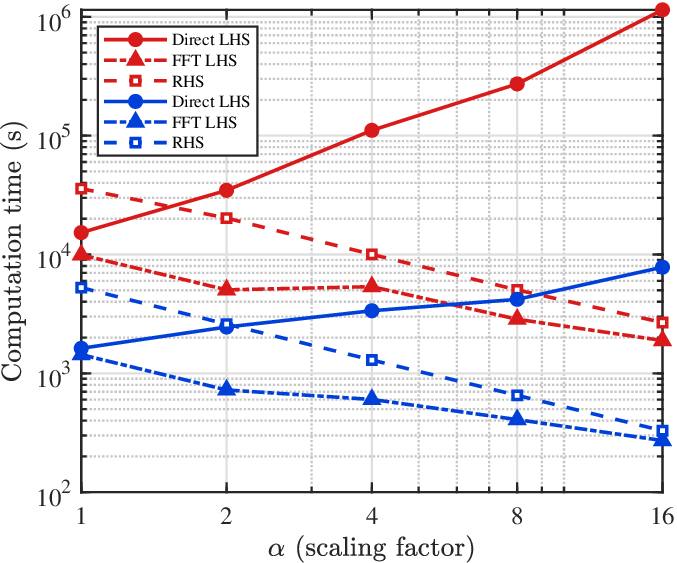}
    \caption{Comparison of the computation times associated with the LHS and RHS terms, defined in \eqref{eq:lhs_rhs_def}, plotted versus time-step scaling factor $\alpha$ for the coarse (blue) and fine voxelizations (red) shown in Table~\ref{tab:voxelization}.}
    \label{fig:Section_V_inhomogeneous_Cube_LHS_RHS_dir_fft}
\end{figure}

The runtime behavior is shown in Fig.~\ref{fig:Section_V_inhomogeneous_Cube_LHS_RHS_dir_fft}. It shows the LHS and RHS computation times for the MOT-TDJVIE system as a function of the time-step scaling factor $\alpha$. As the time-step scaling factor increases from $\alpha=1$ to $\alpha=16$. Increasing $\alpha$ reduces the number of temporal history interactions in the RHS term, lowering its evaluation cost. The RHS contribution is evaluated once per time step, whereas the present-time LHS interaction through $\mathbf{Z}_0$ requires an MVP for every TFQMR iteration for every time step. Consequently, the LHS term dominates in the large-time-step regime. For both voxelizations, replacing the direct LHS operation with FFT-based convolution substantially reduces the overall solver runtime, yielding speedups of up to \(15\times\) at \(\alpha=16\). The effectiveness of the matrix-free FFT evaluation increases as \(\mathbf{Z}_0\) becomes progressively less sparse with increasing time-step size. In the direct sparse-matrix implementation, the cost scales with the number of nonzero entries in \(\mathbf{Z}_0\), whereas the matrix-free FFT-based evaluation scales as \(\mathcal{O}(N\log N)\). Because all simulations are performed in single-threaded mode, the observed runtime reduction is attributed to the matrix-free FFT-based MVP for \(\mathbf{Z}_0\) rather than to parallel hardware acceleration.

\begin{figure}[tb]
    \centering
    \includegraphics[width=0.45\textwidth]{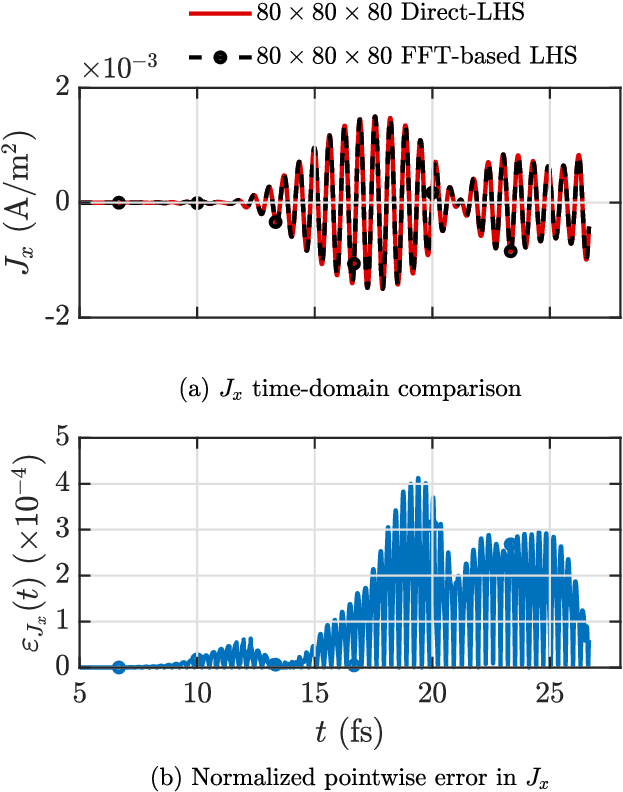}
    \caption{Direct-LHS and FFT-based LHS comparison for the fine mesh $(80\times80\times80)$ with $\alpha = 1$: (a) $J_x$ time-domain response and (b) normalized pointwise error in $J_x$.}
\label{fig:Section_V_J_component_and_error}
\end{figure}

Finally, Fig.~\ref{fig:Section_V_J_component_and_error} verifies the current-density accuracy of the matrix-free FFT-based \(\mathbf{Z}_0\) MVP implementation at the evaluation voxel (41,41,41) of the fine voxel mesh, inside the eight-octant as shown in Fig.~\ref{fig:Section_V_Inhomo_Cube}. Panel (a) compares the direct LHS and the FFT-based LHS time-domain traces of $J_x$ for the fine mesh $(80\times80\times80)$ with $\alpha = 1$, while panel (b) shows the normalized pointwise error in \(J_x\), defined as
\[
\varepsilon_{J_x}(t)
=
\frac{\left|J_x^{\mathrm{FFT}}(t)-J_x^{\mathrm{dir}}(t)\right|}
{\displaystyle \max_t \left|J_x^{\mathrm{dir}}(t)\right|},
\]
where \(J_x^{\mathrm{FFT}}(t)\) denotes the \(x\)-directed current-density response computed using the matrix-free FFT-based implementation of the MOT-JVIE solver, whereas \(J_x^{\mathrm{dir}}(t)\) denotes the corresponding response obtained using the direct implementation. The overlap between the two time-domain traces and the normalized pointwise errors in the order of \(\mathrm{10^{-4}}\) together indicate that the matrix-free FFT-based MVP implementation reproduces the direct MVP solution for the inhomogeneous dielectric case. Overall, the inhomogeneous-cube results demonstrate that the matrix-free FFT-based MVP implementation preserves the convergence behavior and current-density precision of the direct implementation while significantly reducing the dominant LHS computation cost in the large-time-step regime.

\subsection{8$\times$8 Bilayer GST/a-Si Nanopillar Metasurface}

To demonstrate the scalability of the large-time-step MOT-JVIE framework with matrix-free evaluation of the present-time interaction matrix \(\mathbf{Z}_0\), we consider a multiscale dielectric nanopillar metasurface. The geometry is inspired by phase-change-material-based bilayer metasurfaces~\cite{Li2024_PCM_Metasurface}. This example is used as an application-scale computational test case to assess the feasibility of the proposed implementation for large voxelized dielectric structures. The material distribution creates a strongly inhomogeneous volumetric benchmark with subwavelength features, multiple dielectric interfaces, and inter-element coupling. Recent advances in fabricating complex dielectric nanostructures make this class of geometries practically relevant~\cite{Hu2024LaserInduced,Dorrah2025FreeStanding}.

\begin{figure}[b]
    \centering
    \includegraphics[width= 1.05\columnwidth]{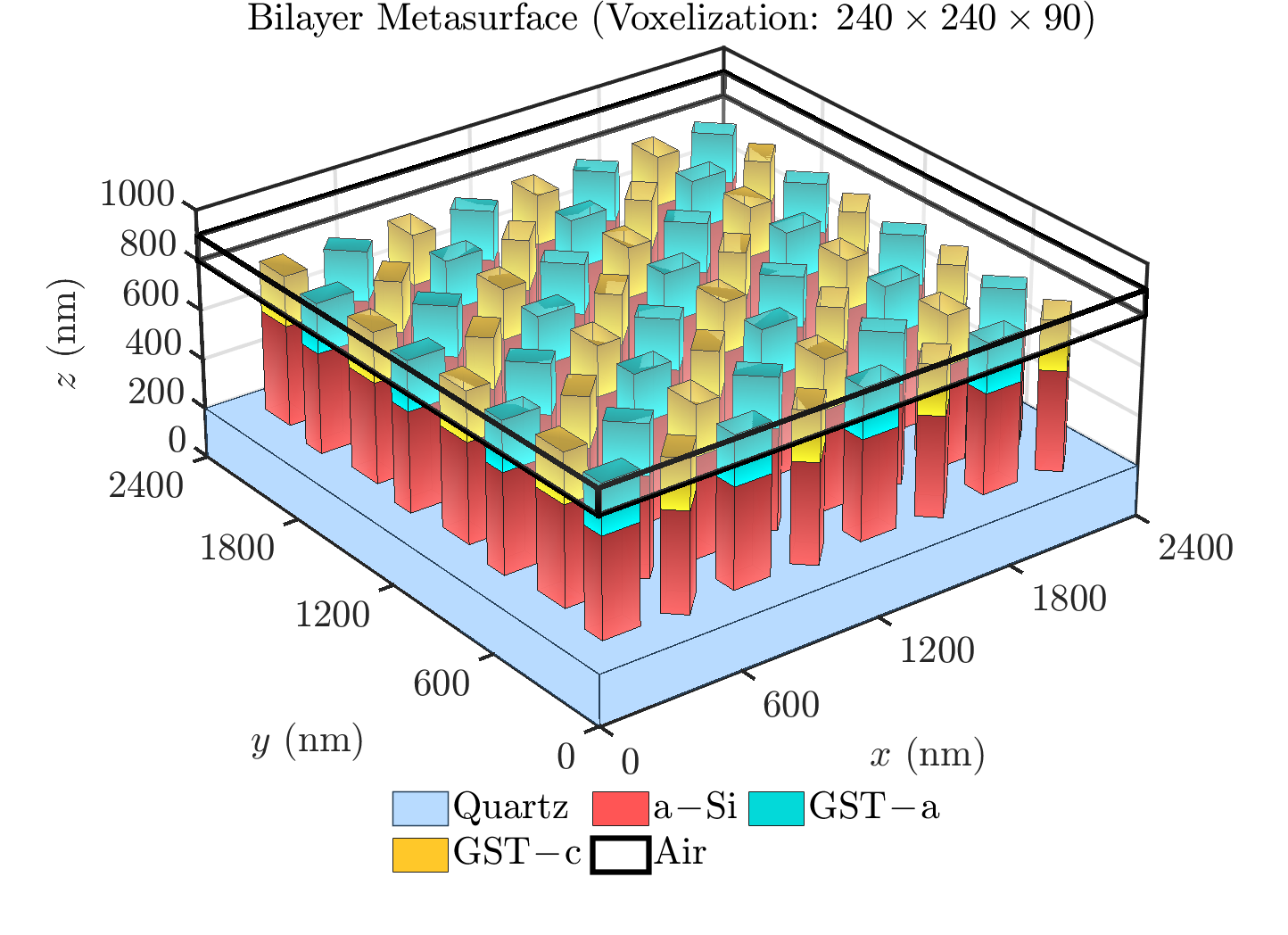}
    \caption{Voxelized geometry of the \(8\times8\) bilayer GST/a-Si nanopillar array. The array is generated by repeating a \(2\times2\) tile with rotation angles \(0^\circ\), \(45^\circ\), \(90^\circ\), and \(135^\circ\).}
    \label{fig:Section_V_8x8_pillars}
\end{figure}

The physical geometry is shown in Fig.~\ref{fig:Section_V_8x8_pillars}. The metasurface consists of $8\times8$ bilayer nanopillars arranged on a regular square lattice. It is generated by repeating a $2\times2$ supercell four times along both transverse directions, with supercell periods $P_x=P_y=560~\mathrm{nm}$. Each supercell contains four nanopillars with rectangular footprint centered at offsets of $\pm 140~\mathrm{nm}$ from the supercell center. The nanopillar footprint has a long edge of $160~\mathrm{nm}$ and a short edge of $100~\mathrm{nm}$. The four pillars have identical dimensions and differ only in their in-plane rotation angles of $0^\circ$, $45^\circ$, $90^\circ$, and $135^\circ$. Each bilayer pillar consists of a $400~\mathrm{nm}$-tall amorphous-silicon (a-Si) base and a $200~\mathrm{nm}$-tall phase-change $\mathrm{Ge_2Sb_2Te_5}$ (GST) cap sharing the same rotated rectangular footprint. Here, GST-a and GST-c denote amorphous GST and crystalline GST, respectively. The array is supported by a $200~\mathrm{nm}$-thick quartz substrate, and a $100~\mathrm{nm}$ air region is included above the metasurface.

The material configuration is shown in Fig.~\ref{fig:Section_V_8x8_pillars}. Each voxel is assigned a scalar, nondispersive relative permittivity according to the material it occupies, thereby isolating the computational scalability of the large-time-step MOT-JVIE framework. Incorporating dispersive, phase-transition, or nonlinear material models is beyond the present scope and remains a subject for future extension. The values used in the simulation are $\varepsilon_{r,\mathrm{air}}=1.0$, $\varepsilon_{r,\mathrm{quartz}}=2.10$, $\varepsilon_{r,\mathrm{aSi}}=12.0$, $\varepsilon_{r,\mathrm{GST-a}}=16.0$, and $\varepsilon_{r,\mathrm{GST-c}}=25.0$. Within each repeated $2\times2$ supercell, the GST caps of the $0^\circ$ and $90^\circ$ pillars are assigned the amorphous phase, while those of the $45^\circ$ and $135^\circ$ pillars are assigned the crystalline phase. This material assignment creates a spatially varying dielectric contrast across the metasurface and voxel-wise inhomogeneous material loading. 

The metasurface is excited by the Gaussian-modulated plane wave defined in \eqref{eq:Einc_modGauss}. The incident field is $x$-polarized and propagates along the $-z$ direction. The pulse is centered at a wavelength of approximately $960~\mathrm{nm}$, corresponding to $f_0=312~\mathrm{THz}$, and its spectral content covers the near-infrared range of approximately $800$--$1100~\mathrm{nm}$. The Gaussian pulse width and delay are $\sigma=11.06~\mathrm{fs}$ and $t_0=68.85~\mathrm{fs}$, respectively. The voxelized computational domain has dimensions $2400~\mathrm{nm}\times2400~\mathrm{nm}\times900~\mathrm{nm}$ and is discretized using a uniform Cartesian mesh with $\Delta x=\Delta y=\Delta z=10~\mathrm{nm}$. This results in a $240\times240\times90$ voxel grid, containing $5.184\times10^6$ cells and approximately $1.56\times10^7$ vector current-density unknowns per time step when all three Cartesian components are retained. For this spatial discretization, the air-based CFL time step is $\Delta t_{\mathrm{CFL}}=0.0193~\mathrm{fs}$, which is used as a reference time-step scale. The MOT-JVIE formulation is not constrained by this CFL limit. Instead, the time step is selected which satisfy the temporal sampling requirement in ~\eqref{eq:dt_nyquist_tdjevie} for the prescribed incident pulse. The maximum frequency that is realistically represented in the modulated Gaussian pulse is $f_{\max}=375~\mathrm{THz}$. Using \eqref{eq:dt_nyquist_tdjevie} with $\kappa=2$ gives a sampling-based limit of $\Delta t_{\mathrm{Nyq}}=0.67~\mathrm{fs}$, corresponding to approximately $34\Delta t_{\mathrm{CFL}}$. The simulation uses $\Delta t=16\Delta t_{\mathrm{CFL}}=0.308~\mathrm{fs}$, which remains below the sampling limit while reducing the number of time steps by a factor of 16 relative to the CFL-based time-step scale on the same spatial grid.

The full metasurface response is simulated over a total time window of $T_{\mathrm{sim}}=385.17~\mathrm{fs}$ using the selected time step $\Delta t=0.308~\mathrm{fs}$. On the single-threaded computational platform specified at the beginning of this section, the matrix-free FFT-based \(\mathbf{Z}_0\) MVP implementation completes the full time-domain simulation in approximately $2$ hours. A direct MOT-JVIE implementation with the same voxelization and time step is not practical at this problem size because explicit sparse storage of the present-time interaction matrix \(\mathbf{Z}_0\) exceeds the available system memory of \(2~\mathrm{TB}\). Thus, this application-scale metasurface example demonstrates full-array scalability of the matrix-free implementation rather than direct-versus-FFT runtime comparison.

\begin{figure}[tb]
    \centering
    \includegraphics[width=1.0\columnwidth]{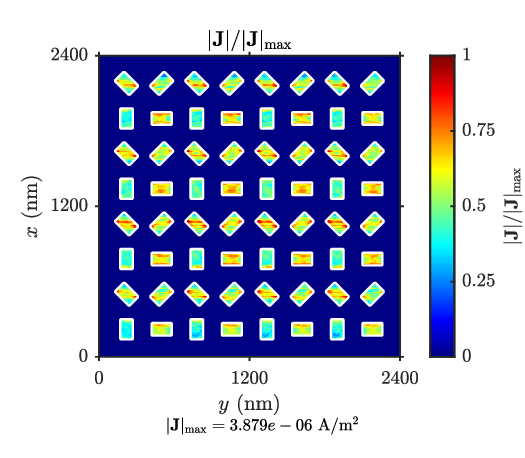}
    \caption{Normalized total current-density magnitude, $|\mathbf{J}|/|\mathbf{J}|_{\max}$, on the $xy$-plane at $z=750~\mathrm{nm}$. The axes are shown in nanometers over the $2400~\mathrm{nm}\times2400~\mathrm{nm}$ simulation domain. White contours denote dielectric-material boundaries.}
    \label{fig:Section_V_8x8_pillars_J_750_nm}
\end{figure}

Fig.~\ref{fig:Section_V_8x8_pillars_J_750_nm} shows the normalized total current-density magnitude, $|\mathbf{J}|/|\mathbf{J}|_{\max}$, on the $xy$-plane at $z=750~\mathrm{nm}$ and $t=308.13~\mathrm{fs}$. This plane intersects the upper GST layer of the bilayer nanopillars, where the material distribution includes both GST-a and GST-c regions. The white contours indicate the dielectric-material boundaries within the full computational domain. The resulting spatial distribution illustrates the induced volumetric current-density response of the complete $8\times8$ metasurface, demonstrating that the FFT-accelerated MOT-TDJVIE implementation can simulate the full voxelized structure without assembling the present-time interaction matrix $\mathbf{Z}_0$.

Overall, this example demonstrates that the large-time-step MOT-JVIE framework with matrix-free \(\mathbf{Z}_0\) evaluation enables time-domain analysis of a multiscale nanopillar metasurface with approximately $1.56\times10^7$ vector current-density unknowns. The matrix-free FFT-based \(\mathbf{Z}_0\) MVP implementation removes the prohibitive memory requirement associated with explicit storage of \(\mathbf{Z}_0\), while the use of $\Delta t=16\Delta t_{\mathrm{CFL}}$ reduces the number of time steps subject to the temporal sampling requirement in~\eqref{eq:dt_nyquist_tdjevie}. Together, these two mechanisms make TDVIE analysis feasible for large volumetric photonic structures that are impractical with a direct MOT-JVIE implementation.


\section{Conclusion}\label{sec:Conclusion}

We investigated large-time-step operation in a marching-on-in-time current-density volume integral equation solver for transient dielectric scattering from two complementary aspects: the feasibility of employing enlarged time steps while preserving the consistency of the computed transient responses, and the computational challenges introduced by this large-time-step regime. In the considered MOT-JVIE formulation, the time-step size is not restricted by the CFL condition associated with the spatial discretization, but must satisfy the temporal sampling requirement of the incident transient signal. For the wavelengths and bandwidths considered here, this enabled accurate responses with time steps up to $16$ times larger than the CFL reference value. Once this sampling requirement was no longer satisfied, temporal under-resolution produced observable phase and amplitude deviations.

The computational analysis further showed that enlarging the time step increases the number of nonzero entries in the present-time interaction matrix $\mathbf{Z}_0$. Since MVPs for $\mathbf{Z}_0$ are required during each linear solve at every time step, this operation becomes the dominant cost in the large-time-step regime. This bottleneck was addressed using a matrix-free FFT-based MVP implementation for $\mathbf{Z}_0$, in which an established Toeplitz/circulant FFT framework is applied to the Green-function part Green-function-related volume-integral operator. The resulting operator application is evaluated through FFT-based convolution, avoiding explicit assembly and storage of $\mathbf{Z}_0$ and reducing the dominant MVP cost to $\mathcal{O}(N\log N)$. Numerical results for dielectric-cube examples and an $8\times8$ bilayer nanopillar metasurface demonstrated the accuracy, efficiency, and scalability of the proposed implementation. In single-threaded execution, the method was demonstrated for a problem with approximately $1.56\times10^7$ vector current-density unknowns, representing, to the best of the authors' knowledge, the first reported single-threaded large-time-step MOT-JVIE demonstration beyond $15$ million unknowns. These results demonstrate the feasibility of the large-time-step MOT-JVIE solver for large voxelized domains and indicate its potential applicability to representative nanophotonic dielectric structures.

\section{Acknowledgment} 
This work was supported by Holland High Tech under Project MetaLenses - 22.0656 and co-financed through a PPS innovation subsidy for public–private research and development collaboration.

\bibliographystyle{IEEEtran}
\bibliography{reference}

@IEEEtranBSTCTL{IEEEexample:BSTcontrol,
  CTLdash_repeated_names = "no"
}

@book{ergul2019new,
  title={New trends in computational electromagnetics},
  editor = {Ozgur Ergul},
  year={2019},
  doi = {10.1049/SBEW533E},
  publisher={SciTech Publishing, Inc.}
}

@article{jin2019multiphysics,
  title={Multiphysics modeling in electromagnetics: Technical challenges and potential solutions},
  author={Jin, Jian-Ming and Yan, Su},
  journal={IEEE Antennas and Propagation Magazine},
  volume={61},
  number={2},
  pages={14--26},
  year={2019},
  doi={10.1109/MAP.2019.2895623},
  publisher={IEEE}
}

@article{zhang2020electromagnetic,
  title={Electromagnetic-circuital-thermal multiphysics simulation method: A review},
  author={Zhang, Huan Huan and Wang, Pan Pan and Zhang, Shuai and Li, Long and Li, Ping and Sha, Wei EI and Jiang, Li Jun},
  journal={Progress In Electromagnetics Research},
  volume={169},
  pages={87--101},
  year={2020},
  doi={10.2528/PIER20112801},
  publisher={EMW Publishing}
}

@book{Taflove1995,
  author    = {Taflove, Allen},
  title     = {Computational Electrodynamics: The Finite-Difference Time-Domain Method},
  publisher = {Artech House},
  address   = {Boston, MA, USA},
  year      = {1995}
}

@article{He2012LargeStepTDFEM,
  author  = {He, Q. and Gan, H. and Jiao, D.},
  title   = {Explicit Time-Domain Finite-Element Method Stabilized for an Arbitrarily Large Time Step},
  journal = {IEEE Transactions on Antennas and Propagation},
  volume  = {60},
  number  = {11},
  pages   = {5240--5250},
  year    = {2012},
  month   = nov,
  doi     = {10.1109/TAP.2012.2207666}
}

@article{Makwana2018LargeStepsEM,
  author  = {Makwana, Nikitabahen N. and Chatterjee, Avijit},
  title   = {Computing with Large Time Steps in Time-Domain Electromagnetics},
  journal = {Journal of Electromagnetic Waves and Applications},
  volume  = {32},
  number  = {17},
  pages   = {2182--2194},
  year    = {2018},
  doi     = {10.1080/09205071.2018.1500314}
}

@article{Makwana2019LargeStepsLayered,
  author  = {Makwana, Nikitabahen Navinchandra and Chatterjee, Avijit},
  title   = {Computing with Large Time Steps for Electromagnetic Wave Propagation in Multilayered Homogeneous Media},
  journal = {Progress In Electromagnetics Research M},
  volume  = {80},
  pages   = {45--56},
  year    = {2019},
  doi     = {10.2528/PIERM19011402}
}

@article{Gaffar2014USFDTD,
  author  = {Gaffar, M. and Jiao, D.},
  title   = {An Explicit and Unconditionally Stable {FDTD} Method for Electromagnetic Analysis},
  journal = {IEEE Transactions on Microwave Theory and Techniques},
  volume  = {62},
  number  = {11},
  pages   = {2538--2550},
  year    = {2014},
  month   = nov,
  doi     = {10.1109/TMTT.2014.2358557}
}

@article{Andriulli2009TDEFIE,
  author  = {Andriulli, Francesco P. and Ba{\u{g}}c{\i}, Hakan and Vipiana, Francesca and Vecchi, Giuseppe and Michielssen, Eric},
  title   = {Analysis and Regularization of the {TD-EFIE} Low-Frequency Breakdown},
  journal = {IEEE Transactions on Antennas and Propagation},
  volume  = {57},
  number  = {7},
  pages   = {2034--2046},
  year    = {2009},
  month   = jul,
  doi     = {10.1109/TAP.2009.2019887}
}

@article{Beghein2015LargeStepTDEFIE,
  author  = {Beghein, Yves and Cools, Kristof and Andriulli, Francesco P.},
  title   = {A {DC} Stable and Large-Time-Step Well-Balanced {TD-EFIE} Based on Quasi-{Helmholtz} Projectors},
  journal = {IEEE Transactions on Antennas and Propagation},
  volume  = {63},
  number  = {7},
  pages   = {3087--3097},
  year    = {2015},
  month   = jul,
  doi     = {10.1109/TAP.2015.2426796}
}

@article{Dely2020IRKTDEFIE,
  author  = {D{\'e}ly, Alexandre and Andriulli, Francesco P. and Cools, Kristof},
  title   = {Large Time Step and {DC} Stable {TD-EFIE} Discretized With Implicit {Runge--Kutta} Methods},
  journal = {IEEE Transactions on Antennas and Propagation},
  volume  = {68},
  number  = {2},
  pages   = {976--985},
  year    = {2019},
  month   = feb,
  doi     = {10.1109/TAP.2019.2943443}
}

@article{Sankaran2019RightTools,
  author  = {Sankaran, Krishnaswamy},
  title   = {Are You Using the Right Tools in Computational Electromagnetics?},
  journal = {Engineering Reports},
  volume  = {1},
  number  = {3},
  pages = {e12041},
  year    = {2019},
  doi     = {10.1002/eng2.12041}
}

@incollection{deHoop2008Reciprocity,
  author    = {de Hoop, Adrianus T.},
  title     = {Electromagnetic Reciprocity Theorems and Their Applications},
  booktitle = {Handbook of Radiation and Scattering of Waves: Acoustic Waves in Fluids, Elastic Waves in Solids, Electromagnetic Waves},
  pages     = {851--857},
  year      = {2008}
}

@article{Gres2001TransientVIE,
  author  = {Gres, Noel T. and Ergin, A. Arif and Michielssen, Eric and Shanker, Balasubramaniam},
  title   = {Volume-Integral-Equation-Based Analysis of Transient Electromagnetic Scattering From Three-Dimensional Inhomogeneous Dielectric Objects},
  journal = {Radio Science},
  volume  = {36},
  number  = {3},
  pages   = {379--386},
  year    = {2001},
  doi     = {10.1029/2000RS002342}
}

@article{Shanker2004FastLossy,
  title={Fast analysis of transient scattering from lossy inhomogeneous dielectric bodies},
  author={Shanker, B and Ayg{\"u}n, K and Michielssen, E},
  journal={Radio Science},
  volume={39},
  number={2},
  pages={1--14},
  year={2004},
  publisher={AGU}
}

@article{Kobidze2005FastTDIEScheme,
  author  = {Kobidze, Gregory and Gao, Jun and Shanker, Balasubramaniam and Michielssen, Eric},
  title   = {A Fast Time Domain Integral Equation Based Scheme for Analyzing Scattering From Dispersive Objects},
  journal = {IEEE Transactions on Antennas and Propagation},
  volume  = {53},
  number  = {3},
  pages   = {1215--1226},
  year    = {2005},
  doi     = {10.1109/TAP.2004.841295}
}

@article{Hu2016NonconformalTDVIE,
  author  = {Hu, Y. L. and Li, J. and Ding, D. Z. and Chen, R. S.},
  title   = {Analysis of Transient EM Scattering From Penetrable Objects by Time Domain Nonconformal VIE},
  journal = {IEEE Transactions on Antennas and Propagation},
  volume  = {64},
  number  = {1},
  pages   = {360--365},
  year    = {2015},
  doi     = {10.1109/TAP.2015.2501437}
}

@article{Cao2017HighOrderNystromTDVIE,
  author  = {Cao, Jun and Ding, Dazhi and Cheng, Guangshang and Chen, Rushan},
  title   = {A Higher Order {Nystr{\"o}m} {TD-VIE} Method for Scattering From Magnetized Plasma Objects},
  journal = {IEEE Antennas and Wireless Propagation Letters},
  volume  = {16},
  pages   = {408--411},
  year    = {2016},
  doi     = {10.1109/LAWP.2016.2641019}
}

@article{Sayed2015,
  author  = {Sayed, Sadeed Bin and {\"U}lk{\"u}, H{\"u}seyin Arda and Ba{\u{g}}c{\i}, Hakan},
  title   = {A Stable Marching-On-in-Time Scheme for Solving the Time-Domain Electric Field Volume Integral Equation on High-Contrast Scatterers},
  journal = {IEEE Transactions on Antennas and Propagation},
  volume  = {63},
  number  = {7},
  pages   = {3098--3110},
  year    = {2015},
  doi     = {10.1109/TAP.2015.2429736}
}

@article{Sayed2020ExplicitMFVIE,
  author  = {Sayed, Sadeed Bin and {\"U}lk{\"u}, H{\"u}seyin Arda and Ba{\u{g}}c{\i}, Hakan},
  title   = {Explicit Time Marching Schemes for Solving the Magnetic Field Volume Integral Equation},
  journal = {IEEE Transactions on Antennas and Propagation},
  volume  = {68},
  number  = {3},
  pages   = {2224--2237},
  year    = {2019},
  doi     = {10.1109/TAP.2019.2949381}
}

@article{VanDiepen2024PIERB104,
  author  = {van Diepen, Petrus W. N. and van Beurden, Martijn C. and Dilz, Roeland J.},
  title   = {The Influence of Contrast and Temporal Expansion on the Marching-On-in-Time Contrast Current Density Volume Integral Equation},
  journal = {Progress In Electromagnetics Research B},
  volume  = {104},
  year    = {2024},
  doi     = {10.2528/PIERB23091305}
}

@article{Freund1993TFQMR,
  author  = {Freund, Roland W.},
  title   = {A Transpose-Free Quasi-Minimal Residual Algorithm for Non-Hermitian Linear Systems},
  journal = {SIAM Journal on Scientific Computing},
  volume  = {14},
  number  = {2},
  pages   = {470--482},
  year    = {1993},
  doi     = {10.1137/0914029}
}

@article{VanDiepen2024PIERB106,
  author  = {van Diepen, Petrus W. N. and van Beurden, Martijn C. and Dilz, Roeland J.},
  title   = {{FFT}-Acceleration and Stabilization of the {3D} Marching-On-in-Time Contrast Current Density Volume Integral Equation for Scattering From High Contrast Dielectrics},
  journal = {Progress In Electromagnetics Research B},
  volume  = {106},
  pages   = {113--129},
  year    = {2024},
  publisher={Electromagnetics Academy},
  doi     = {10.2528/PIERB24031903}
}

@article{Shannon1949,
    author = {{Shannon}, C.~E.},
    title = "{Communication In The Presence Of Noise}",
    journal = {IEEE Proceedings},
    year = 1998,
    month = feb,
    volume = {86},
    number = {2},
    pages = {447-457},
    doi = {10.1109/JPROC.1998.659497}
}

@article{Unser2000Sampling,
  author  = {Unser, Michael},
  title   = {Sampling---50 Years After {Shannon}},
  journal = {Proceedings of the IEEE},
  volume  = {88},
  number  = {4},
  pages   = {569--587},
  year    = {2000},
  doi     = {10.1109/5.843002}
}

@article{Ergin1999,
  author  = {Ergin, A. Arif and Shanker, Balasubramaniam and Michielssen, Eric},
  title   = {The Plane-Wave Time-Domain Algorithm for the Fast Analysis of Transient Wave Phenomena},
  journal = {IEEE Antennas and Propagation Magazine},
  volume  = {41},
  number  = {4},
  pages   = {39--52},
  year    = {1999},
  doi     = {10.1109/74.789736}
}

@article{Liu2016PWTDAcceleratedExplicitEFVIE,
  author  = {Liu, Yang and Al-Jarro, Ahmed and Ba{\u{g}}c{\i}, Hakan and Michielssen, Eric},
  title   = {Parallel {PWTD}-Accelerated Explicit Solution of the Time-Domain Electric Field Volume Integral Equation},
  journal = {IEEE Transactions on Antennas and Propagation},
  volume  = {64},
  number  = {6},
  pages   = {2378--2388},
  year    = {2016},
  doi     = {10.1109/TAP.2016.2546964}
}

@article{Yilmaz2004,
  author  = {Yilmaz, Ali E. and Jin, Jian-Ming and Michielssen, Eric},
  title   = {Time Domain Adaptive Integral Method for Surface Integral Equations},
  journal = {IEEE Transactions on Antennas and Propagation},
  volume  = {52},
  number  = {10},
  pages   = {2692--2708},
  year    = {2004},
  doi     = {10.1109/TAP.2004.834399}
}

@article{Shi2011,
  author  = {Shi, Yan and Jin, Jian-Ming},
  title   = {A Time-Domain Volume Integral Equation and Its Marching-On-in-Degree Solution for Analysis of Dispersive Dielectric Objects},
  journal = {IEEE Transactions on Antennas and Propagation},
  volume  = {59},
  number  = {3},
  pages   = {969--978},
  year    = {2010},
  doi     = {10.1109/TAP.2010.2103038}
}

@article{Zwamborn1991,
  author  = {Zwamborn, A. P. M. and van den Berg, P. M.},
  title   = {A Weak Form of the Conjugate Gradient {FFT} Method for Two-Dimensional {TE} Scattering Problems},
  journal = {IEEE Transactions on Microwave Theory and Techniques},
  volume  = {39},
  number  = {6},
  pages   = {953--960},
  year    = {1991},
  doi     = {10.1109/22.81664}
}

@article{Zwamborn1992,
  author  = {Zwamborn, P. and van den Berg, P. M.},
  title   = {The Three-Dimensional Weak Form of the Conjugate Gradient {FFT} Method for Solving Scattering Problems},
  journal = {IEEE Transactions on Microwave Theory and Techniques},
  volume  = {40},
  number  = {9},
  pages   = {1757--1766},
  year    = {1992},
  doi     = {10.1109/22.156602}
}

@article{Polimeridis2014StableFFTJVIE,
  author  = {Polimeridis, Athanasios G. and Villena, Juan F. and Daniel, Luca and White, Jacob K.},
  title   = {Stable {FFT-JVIE} Solvers for Fast Analysis of Highly Inhomogeneous Dielectric Objects},
  journal = {Journal of Computational Physics},
  volume  = {269},
  pages   = {280--296},
  year    = {2014},
  publisher={Elsevier},
  doi     = {10.1016/j.jcp.2014.03.026}
}

@article{Yilmaz2002,
  author  = {Yilmaz, Ali E. and Weile, Daniel S. and Jin, Han-Ming and Michielssen, Eric},
  title   = {A Hierarchical {FFT} Algorithm ({HIL-FFT}) for the Fast Analysis of Transient Electromagnetic Scattering Phenomena},
  journal = {IEEE Transactions on Antennas and Propagation},
  volume  = {50},
  number  = {7},
  pages   = {971--982},
  year    = {2002},
  publisher={IEEE},
  doi     = {10.1109/TAP.2002.802094}
}

@phdthesis{Yilmaz2005,
  author = {Yilmaz, Ali E.},
  title  = {Parallel {FFT}-Accelerated Time-Domain Integral Equation Solvers for Electromagnetic Analysis},
  school = {University of Illinois at Urbana-Champaign},
  address = {Urbana, IL, USA},
  year   = {2005}
}

@article{ChanOlkin1994ToeplitzBlock,
  author  = {Chan, Tony F. and Olkin, Julia A.},
  title   = {Circulant Preconditioners for Toeplitz-Block Matrices},
  journal = {Numerical Algorithms},
  volume  = {6},
  number  = {1},
  pages   = {89--101},
  year    = {1994},
  doi     = {10.1007/BF02149764}
}

@article{Li2024_PCM_Metasurface,
  author  = {Li, Chensheng and Du, Shuo and Pan, Ruhao and Xiong, Xiaoyu and Tang, Zhiyang and Zheng, Ruixuan and Liu, Yunan and Geng, Guangzhou and Sun, Jingbo and Gu, Changzhi and Guo, Haiming and Li, Junjie},
  title   = {Phase Change Materials-Based Bilayer Metasurfaces for Near-Infrared Photonic Routing},
  journal = {Advanced Functional Materials},
  volume  = {34},
  number  = {14},
  pages   = {2310626},
  year    = {2024},
  doi     = {10.1002/adfm.202310626}
}

@article{Hu2024LaserInduced,
  author  = {Hu, Sha and Wang, Chao and Du, Shuo and Han, Zhanghua and Hu, Ning and Gu, Changzhi},
  title   = {Laser-induced reconfigurable wavefront control with a structured {Ge2Sb2Te5}-based metasurface},
  journal = {Communications Physics},
  volume  = {7},
  number={1},
  pages   = {346},
  year    = {2024},
  publisher={Nature Publishing Group UK London},
  doi     = {10.1038/s42005-024-01846-9}
}

@article{Dorrah2025FreeStanding,
  author  = {Dorrah, Ahmed H. and Park, Joon-Suh and Palmieri, Alfonso and Capasso, Federico},
  title   = {Free-standing bilayer metasurfaces in the visible},
  journal = {Nature Communications},
  volume  = {16},
  number={1},
  pages   = {3126},
  year    = {2025},
  publisher={Nature Publishing Group UK London},
  doi     = {10.1038/s41467-025-58205-7}
}

\end{document}